\documentclass[aps,nofootinbib,preprint]{revtex4}
\usepackage{amsfonts,amssymb,amsmath}
\usepackage{enumerate}

\newcommand{\beq}{\begin{equation}}
\newcommand{\eeq}{\end{equation}}
\newcommand{\beqa}{\begin{eqnarray}}
\newcommand{\eeqa}{\end{eqnarray}}
\newcommand{\bfx}{{\bf x}}
\newcommand{\bfk}{{\bf k}}

\begin{document}

\title{Analytic solution for matter density perturbations
in a class of viable cosmological $f(R)$ models}

\author{Hayato Motohashi\footnote{motohashi@resceu.s.u-tokyo.ac.jp} $^{~1,2}$,
Alexei A. Starobinsky\footnote{alstar@landau.ac.ru} $^{~2,3}$ and
Jun'ichi Yokoyama\footnote{yokoyama@resceu.s.u-tokyo.ac.jp}$^{~2,4}$}
\address{$^{1}$ Department of Physics, Graduate School of Science,\\
The University of Tokyo, Tokyo 113-0033, Japan \\
$^{2}$ Research Center for the Early Universe (RESCEU), 
Graduate School of Science, The University of Tokyo, Tokyo 113-0033, Japan \\
$^{3}$ L. D. Landau Institute for Theoretical Physics, 
Moscow 119334, Russia \\
$^{4}$ Institute for the Physics and Mathematics of the Universe(IPMU), \\
The University of Tokyo, Kashiwa, Chiba, 277-8568, Japan}

\begin{abstract}
For a class of viable cosmological models in $f(R)$
gravity in which deviation from the Einstein gravity decreases 
as a inverse power law of the Ricci
scalar $R$ for large $R$, an analytic solution for density
perturbations in the matter component during the matter dominated stage
is obtained in terms of hypergeometric functions. An analytical expression for the
matter transfer function at scales much less than the present Hubble
scale is also obtained.
\end{abstract}

\begin{flushright}
RESCEU-23/09
\end{flushright}

\maketitle

\section{Introduction}

Recent numerous observational data obtained from angular
anisotropy and polarization of the cosmic microwave background
radiation, large-scale gravitational clustering of galaxies and
observations of distant supernovae explosions prove convincingly
that the Universe expands with acceleration at the present time
while it was decelerating in the past for redshifts larger than
about $0.7$. If interpreted in terms of the Einstein general
theory of relativity, this acceleration requires the existence of
some new component in the right-hand side of the Einstein
equations, dubbed dark energy (DE), which remains practically
unclustered at all scales at which gravitational clustering of
baryonic and dark non-baryonic matter is seen, and which effective
pressure $p_{DE}$ is approximately equal to minus its effective
energy density $\rho_{DE}$. Thus, its properties are very close to
those of a cosmological constant $\Lambda$ (see Refs.~
\cite{SS00}~-~\cite{FTH08} for some reviews). The simplest
possibility of DE being exactly $\Lambda$ combined with a
non-relativistic non-baryonic dark matter (the standard spatially
flat $\Lambda$CDM cosmological model) provides a good fit to all
existing observational data\cite{Komatsu:2008hk}. In this case
$\Lambda$ acquires the status of a new fundamental physical
constant. However, its required value is very small as compared to
known atomic and elementary particle scales (not speaking about
the Planck ones), so a firm theoretical prediction for this
quantity from first principles is lacking currently (although it
may arise due to some non-perturbative effects, see e.g.
Ref.~\cite{Yokoyama:2001ez}).

On the other hand, in the second case when a component with
qualitatively similar properties is assumed to exist -- in the
inflationary scenario of the early Universe, we are sure that this
``primordial DE" may not be an exact cosmological constant since it
should decay long time ago. That is why it is natural to assume by
analogy that the present DE is not absolutely stable, too.

An interesting alternative to the standard $\Lambda$CDM model is
provided by ``geometric" DE models based on $f(R)$ gravity which
modify and generalize the Einstein General Relativity by
introducing a new function of Ricci scalar, $f(R)$, into the
gravitational field action instead of $R-2\Lambda$ (see
Ref.~\cite{SF08} for a recent review). $f(R)$ gravity, 
which in turn is
a particular case of more general scalar-tensor gravity, contains
an additional scalar degree of freedom or, in quantum language, a
massive scalar particle.
\footnote{This particle was dubbed 
"scalaron" in Ref.~\cite{S80} where the particular case $f(R)=
R+R^2/6M^2$ (plus small additional terms) was used to construct
the first internally self-consistent inflationary scenario of the
early Universe having a graceful exit to the subsequent
radiation-dominated Friedmann-Lema\^\i tre-Robertson-Walker (FLRW)
stage through an intermediate matter-dominated reheating period.
This inflationary model still remains viable since it predicts the
slope of the primordial spectrum of scalar perturbations $n_s$ and
the tensor-to-scalar ratio $r$ in agreement with the most recent
observational data.} 
Because of this, viable models of present DE
in $f(R)$ gravity should satisfy a number of conditions which
exclude many possible, in principle, forms of $f(R)$. In
particular, in order to have the correct Newtonian limit for $R\gg
R(t_0)\sim H_0^2$ where $t_0$ is the present moment and $H_0$ is
the Hubble constant, as well as the standard matter-dominated FLRW
stage with the scale factor behaviour $a(t)\propto t^{2/3}$ driven
by cold dark matter and baryons, the following conditions should
be fulfilled: 
\beq |f(R)-R|\ll R,~~|f'(R)-1|\ll 1,~~Rf''(R)\ll 1, ~~R\gg R_0~, \label{ineq} \eeq 
where the prime denotes the derivative with respect to the argument $R$. 
In addition, the stability condition $f''(R)>0$ has to be satisfied that guarantees
that the standard matter-dominated FLRW stage remains an attractor
with respect to an open set of neighboring isotropic cosmological
solutions in $f(R)$ gravity (in quantum language, this condition
means that scalaron is not a tachyon).\footnote{The second
stability condition $f'(R)>0$ which means that gravity is
attractive and graviton is not a ghost is automatically fulfilled
in this regime.}

A number of functional forms has been proposed that can account
for accelerated expansion without $\Lambda$ while passing
laboratory and astronomical tests\cite{Hu:2007nk,Starobinsky:2007hu}.

The present paper considers the evolution of density perturbation
in $f(R)$ model. In Section 2, we review the $\Lambda$CDM model.
In Section 3, after discussing density perturbation in general
$f(R)$ gravity, we choose a specific inverse power-law form of
$f(R)-R$ which is the limiting form of the models proposed in
Refs.~\cite{Hu:2007nk,Starobinsky:2007hu} for $R\gg R_0$ and find an
analytic solution for density perturbations.

\section{Density perturbations in the $\Lambda$CDM model}

Perturbed spatially flat FLRW metric in the longitudinal gauge is
\beq ds^2=-(1+2\Phi)dt^2+a^2(t)(1-2\Psi) (dr^2+r^2d\theta+r^2\sin^2\theta d\phi). \eeq 
During the
matter-dominated era, non-zero components of the energy-momentum
tensor are given by 
\beq \label{matd} T^0_0=-\rho-\delta \rho,\quad T^0_i=-\rho \partial_i v , \eeq 
where $\rho$ and $\delta \rho$ denote
background matter density and its fluctuation, and $v$ is the
velocity potential of scalar perturbation.

From the Einstein equations, we can derive the differential equation
for comoving density perturbation defined by
\beq \delta = \frac{\delta \rho}{\rho} + 3Hav. \eeq
Fourier transformation of density perturbation is given by
\beq \delta_{\bfk}(t) = \int \frac{d^3x}{(2\pi )^{3/2}} \delta (t, \bfx) e^{i\bfk \cdot \bfx}, \eeq
where $\bfk$ denotes comoving wavenumber.
In the following, we abbreviate $\delta_\bfk(t)$ just $\delta$
for simplicity.
The evolution of $\delta$ in Fourier space is governed by
\beq  \ddot \delta + 2H \dot \delta - 4\pi G \rho \delta = 0. \eeq
In matter dominated era, this equation becomes
\beq \label{de0} \ddot \delta + \frac{4}{3t} \dot \delta - \frac{2}{3t^2} \delta = 0. \eeq
While it has two independent solutions proportional to $t^{2/3}$
and $t^{-1}$,
we are only interested in the growing mode:
\beq \label{solLCDM} \delta_{\bfk}(t)= \delta_{0\bfk} \left(\frac{t}{t_0}\right)^{\frac{2}{3}}. \eeq
Here, $\delta_{0\bfk}$ is the initial value $\delta_{\bfk}(t_0)$.

\section{Density perturbations in $f(R)$ gravity}

We write the action in the following form:
\beq \label{ac} S=\int d^4x \sqrt{-g}
\left[ \frac{1}{16\pi G}f(R) + \mathcal{L}^{(m)} \right], \eeq
where $G$ is the Newton constant and $\mathcal{L}^{(m)}$ is matter Lagrangian.
If we take $f(R)=R-2\Lambda$, Eq.~\eqref{ac} reduces to the
Einstein-Hilbert action for the $\Lambda$CDM model.
Below we consider $f(R)$ which vanishes for $R=0$, so no cosmological
constant is introduced by hand in flat space-time.

In $f(R)$ gravity, modified Einstein equations have the form: 
\beq FR_{\mu\nu}-\frac{1}{2}fg_{\mu\nu}+(g_{\mu\nu}\square-\nabla_\mu \nabla_\nu )F=8\pi G T_{\mu\nu}, \eeq 
where 
\beq F(R)\equiv\frac{df}{dR}. \eeq 
For dust-like matter \eqref{matd}, the background
equations take the form: \beqa
3FH^2&=&\frac{1}{2}(FR-f)-3H\dot F+ 8\pi G \rho, \\
-2F\dot H&=&\ddot F-H\dot F+ 8\pi G\rho, \\
\dot \rho+3H\rho&=&0.
\eeqa
When deviation from the Einstein gravity is small, namely,
$f(R)\cong R$ and $F(R)\cong 1$, these equations yield the standard
matter-dominated regime $a(t)=a_0(t/t_0)^{2/3}$.

The differential equation describing a density perturbation in the
subhorizon regime is: 
\beq \label{de1} \ddot \delta + 2H\dot \delta - 4\pi G_{\text{eff}}\rho \delta = 0, \eeq 
where 
\beqa
\label{tsuji} G_{\text{eff}}&=&\frac{G}{F}
\frac{1+4\frac{k^2}{a^2}\frac{F_{,R}}{F}}
{1+3\frac{k^2}{a^2}\frac{F_{,R}}{F}}.
\eeqa 
The above equation was derived by
Tsujikawa\cite{Tsujikawa:2007gd} (see also Ref.~\cite{SHS07}) who took
the subhorizon limit of the coupled equations of $\delta$ and
metric perturbations before reducing the system to a single
equation.\footnote{
In the two opposite limits
$\frac{k^2}{a^2}\frac{F_{,R}}{F}\ll 1$ and
$\frac{k^2}{a^2}\frac{F_{,R}}{F}\gg 1$, Eq.~\eqref{de1} reduces to
the one derived in Ref.~\cite{BEPS00} for matter density perturbations
in general scalar-tensor cosmology for the values of the
Brans-Dicke parameter $\omega_{BD}\gg 1$ and $\omega_{BD}=0$
correspondingly.} 
Its solutions were numerically studied in 
Refs.~\cite{SPH07}~-~\cite{GMP09} in the linear regime and
in Refs.~\cite{TT08}~-~\cite{SLOH09} in the non-linear one.

On the other hand, de la Cruz-Dombriz 
{\it et al.}\cite{delaCruzDombriz:2008cp} derived 
the evolution equation of
$\delta$ without any approximation (see also 
Refs.~\cite{ACD08,ACD08a}).
They then took the subhorizon limit to yield Eq.~(33) in 
Ref.~\cite{delaCruzDombriz:2008cp}. We find that, at the
matter-dominated stage, their result can be rewritten in the form
of Eq.~\eqref{de1} but with \beq \label{dela}
G_{\text{eff}}=\frac{G}{F} \frac{F+32\left(
\frac{k^2}{a^2}\frac{F_{,R}}{F} \right)^4} {1+24\left(
\frac{k^2}{a^2}\frac{F_{,R}}{F}\right)^4}. \eeq

Apparently, Eq.~\eqref{tsuji} and Eq.~\eqref{dela} have different
dependence on $k$. However, we can see that both of these two
equations give practically the same result for observationally
relevant $f(R)$ models as follows.

First, we consider two extreme cases with $F_{,R}/F \gg a^2/k^2$
and $F_{,R}/F \ll a^2/k^2$.  In the former case we see both
equations have the same $G_{\text{eff}}$. In the latter case, Eq.~
\eqref{tsuji} gives $G_{\text{eff}}=G/F$, while Eq.~\eqref{dela}
reduces to $G_{\text{eff}}=G$.  However, since we are assuming the
subhorizon regime, we find in this case $F_{,R}/F \ll a^2/k^2 \ll
H^{-2}$, so that $F$ is practically constant (and in fact equal to
unity) at the cosmological time scale.  So, both equations give
the same result in the two extreme regimes.

Next we consider the intermediate regime with $F_{,R}/F \sim
a^2/k^2$.  In this case, Eq.~\eqref{tsuji} and Eq.~\eqref{dela} do
have different wavenumber dependence, but we still have $F_{,R}/F
\sim a^2/k^2 \ll H^{-2}$ and again $F\cong 1$ in cosmological time
scale. It has been numerically confirmed in 
Ref.~\cite{delaCruzDombriz:2008cp} that in such observationally viable
$f(R)$ models density fluctuations evolve in the same way
whichever equations one uses.  Hence both equations give the same
result in this case, too.

We therefore continue to use Eq.~\eqref{de1} with
Eq.~\eqref{tsuji}. From now on, we adopt a specific $f(R)$ model
such that \beq F(R)\equiv
f'(R)=1-\left(\frac{R_0}{R}\right)^{n+1}, \quad n>-1
\label{model}\eeq with $R_0\sim H_0^2$ (but still $R_0<R(t_0)$).
This $f(R)$ corresponds to the models in~
\cite{Hu:2007nk,Starobinsky:2007hu} in the regime $R \gg R_0$, and
we shall use it in this regime only.\footnote{
The parameter $n$
used here has the same sense as in Ref.~\cite{Hu:2007nk}, while it is
equal to $2n$ in the notation of Ref.~\cite{Starobinsky:2007hu}.}
Equation \eqref{de1} then becomes \beq \label{de2} \ddot \delta +
\frac{4}{3t}\dot \delta -
\frac{2}{3t^2}\frac{1+4A(t/t_0)^{2n+8/3}}{1+3A(t/t_0)^{2n+8/3}}
\delta = 0, \eeq with \beq A(n,k)=\frac{(n+1)k^2}{a_0^2R_0^2}.
\eeq

\subsection{Asymptotic behavior}

We can read asymptotic behavior from the differential equation
Eq.~\eqref{de2}.

\begin{enumerate}[(i)]
\item $t\to 0$

In this limit, we can neglect $A(t/t_0)^{2n+8/3}$ respect to $1$.
Therefore, Eq.~\eqref{de2} reduces to Eq.~\eqref{de0}. This
reproduces the result of the $\Lambda$CDM model. The solution is
Eq.~\eqref{solLCDM}, 
\beq \label{t0} \delta_{\bfk}(t) =
\delta_{0\bfk}\left(\frac{t}{t_0}\right)^{\frac{2}{3}}. \eeq

\item $t\to \infty$

Then Eq.~\eqref{de2} reduces 
\beq \ddot \delta + \frac{4}{3t}\dot \delta - \frac{8}{9t^2}\delta = 0. \eeq 
In this regime the growing
mode is given by 
\beq \label{tin} \delta_{\bfk}(t)=\delta_{0\bfk}C(k)\left(\frac{t}{t_0}\right)^{\frac{-1+\sqrt{33}}{6}}. \eeq 
The coefficient $C(k)$ is the transfer function for matter
perturbations which will be derived from the analytic solution found below.
\end{enumerate}

\subsection{Analytic solution}

By changing the variable from $t$ to $\tau =(t/t_0)^\alpha$ with
$\alpha=2n+8/3$, Eq.~\eqref{de2} can be rewritten as 
\beq \delta''+ \left(1+\frac{1}{3\alpha}\right)\frac{\delta'}{\tau} -
\frac{2}{3\alpha^2}\frac{1+4A\tau}{1+3A\tau}\frac{\delta}{\tau^2}=0. \eeq 
Here a prime denotes derivative respect to $\tau$. This
equation can be solved in terms of the hypergeometric function
$_2F_1$. The two independent solutions are 
\beq \label{as}
\delta_{\bfk}(t)=\delta_{0\bfk}\left(\frac{t}{t_0}\right)^{\frac{-1\pm
5}{6}} \,_2F_1\left(\frac{\pm 5-\sqrt{33}}{6\alpha}, \frac{\pm
5+\sqrt{33}}{6\alpha};1\pm\frac{5}{3\alpha};-3A(n,k)\left(\frac{t}{t_0}\right)^\alpha
\right). \eeq 
In the following discussion, we consider the upper
sign case only, because the other solution corresponds to the
decaying mode of perturbations and is singular at $t\to 0$.

Let us check the asymptotic behavior of the solution, Eq.~\eqref{as}.

\begin{enumerate}[(i)]
\item $t\to 0$

\beq \delta_{\bfk}(t) \to \delta_{0\bfk} \left(\frac{t}{t_0}\right)^{\frac{2}{3}}. \eeq

\item $t\to \infty$

\beq \label{tinf} \delta_{\bfk}(t) \to \delta_{0\bfk}
\times \frac{\Gamma\left(1+\frac{5}{3\alpha}\right) \Gamma\left(\frac{\sqrt{33}}
{3\alpha}\right)}{\Gamma\left(1+\frac{5+\sqrt{33}}{6\alpha}\right)
\Gamma\left(\frac{5+\sqrt{33}}{6\alpha}\right)} \left[ \frac{3(n+1)k^2}{a_0^2R_0}\right]
^{\frac{-5+\sqrt{33}}{6\alpha}} \left(\frac{t}{t_0}\right)^{\frac{-1+\sqrt{33}}{6}}. \eeq

\end{enumerate}

In both limits, the asymptotic behavior agrees with that one given
by Eq.~\eqref{t0} and Eq.~\eqref{tin}, respectively. Furthermore,
here we can read off the transfer function, $C(k)$, which appears
in Eq.~\eqref{tin}: \beq
C(k)=\frac{\Gamma\left(1+\frac{5}{2(3n+4)}\right)
\Gamma\left(\frac{\sqrt{33}}{2(3n+4)}\right)}{\Gamma\left(1+\frac{5+\sqrt{33}}{4(3n+4)}\right)
\Gamma\left(\frac{5+\sqrt{33}}{4(3n+4)}\right)} \left[
\frac{3(n+1)k^2}{a_0^2R_0}\right]^{\frac{-5+\sqrt{33}}{4(3n+4)}}.
\label{Ck}\eeq

\section{Conclusions and discussion}

We have obtained an analytic solution describing the growth of
density perturbation at the matter-dominated stage for a specific
class of viable cosmological models in $f(R)$ gravity. Initially,
the solution behaves in the same way as in the $\Lambda$CDM model,
while it experiences an anomalous growth at late times (redshifts
of the order of a few). We also find an analytic expression for
the matter transfer function which shows that an initial
perturbation power spectrum acquires the additional power-law
factor $\propto k^{\Delta n_s}$ with \beq \Delta n_s =
\frac{-5+\sqrt{33}}{3n+4} \label{ns}\eeq at scales much less than
the present Hubble scale, as originally shown in
Ref.~\cite{Starobinsky:2007hu}.\footnote{Note the difference
in the notation for $n$ between Ref.~\cite{Starobinsky:2007hu} 
and our paper which was mentioned above.} 
Clearly, this additional factor
is absent in the matter power spectrum at the recombination time.
So, by comparing the form of the primordial matter power spectrum
derived from CMB data and from galaxy surveys separately, it is
possible to obtain an important constraint on the parameter $n$
characterizing this class of cosmological models in $f(R)$
gravity, although we do not have much stringent constraints on it
at present\cite{Tegmark:2006az}. If we take an upper limit on
$\Delta n_s$ as $\Delta n_s^{\max}=0.05$, which is a conservative
bound for now\cite{Starobinsky:2007hu}, and assume that $R_0$ is
not much less than $H_0^2$ (if otherwise $R_0\ll H_0^2$, deviation
of the background FLRW model from the $\Lambda$CDM one is very
small), we obtain a constraint 
\beq   n > 4.96\left(\frac{\Delta n_s^{\max}}{0.05}\right)^{-1}-1.33. \eeq 
Future observational data together with a more detailed
theoretical analysis may well yield a more stringent bound on $n$.

Of course, the $f(R)$ gravity model \eqref{model} is viable for a
finite range of $R$ only, in particular, for $R\gg R_0$. For
$R\sim R(t_0)\sim H_0^2$, it has to be substituted by a more
complicated expression admitting a stable (or, at least
metastable) de Sitter solution, e.g. by the models presented in 
Refs.~\cite{Hu:2007nk,Starobinsky:2007hu}. 
As a result, the equation for
matter density perturbations has to be solved numerically for
recent redshifts $z\lesssim 1$. However, evolution in this region
may add only a $k$-independent factor to the total matter transfer
function. Therefore, the $k$ exponent in Eqs.~\eqref{Ck} and \eqref{ns}
does not depend on a concrete form of $f(R)$ for $R\sim R(t_0)$.

Also, the model \eqref{model} should not be used for too large
values of $R$ for several reasons. First, the effective scalaron
mass squared $m_s^2= 1/3F_{,R}$ (in the regime $F_{,R}H^2\ll 1$) grows
quickly with $R$ to the past and may even exceed the Planck mass
making copious production of primordial black holes possible. As
was noted in Ref.~\cite{Starobinsky:2007hu}, this problem may be
avoided by adding the $R^2/6M^2$ term to $f(R)$ that bounds the
scalaron mass from above just by $M$. Simultaneously, such a
change of $f(R)$ at large $R$ removes the ``Big Boost'' singularity
(in terminology of Ref.~\cite{BDK08} where such a singularity appeared
in a different context), which generic appearance in the model
\eqref{model} was shown in Ref.~\cite{F08}. However, the value of $M$
should be sufficiently large in order not to destroy the standard
cosmology of the present and early Universe. In particular, its
values considered in Refs.~\cite{D08,KM09} seem not to be
high enough for this purpose. Indeed, the simplest way to solve
one more problem of this $f(R)$ cosmological scenario (also noted
in Ref.~\cite{Starobinsky:2007hu}) -- overproduction of scalarons in
the early Universe -- is, as usual, to have an inflationary
stage preceding the radiation-dominated one. Then $M$ should not be
 smaller than $H$ during its last 60 e-folds, and for $M\approx
3\times 10^{13}$ GeV the scalaron will be the inflaton itself
according to the scenario\cite{S80}.

Note that for sufficiently large $n$ the corresponding correction
to $F(R)$ may become more important than the second term in the
right-hand side of Eq.~\eqref{model} already at the
matter-dominated stage. For example, even for $M$ as large as $3\times
10^{13}$ GeV, the term $R/3M^2$ becomes larger than
$(R_0/R)^{n+1}$ for $R=3\times 10^{10}R(t_0)$ (corresponding to
the matter-radiation equality) if $n\ge 10$. However, this does
not affect the exact solution obtained in the previous section
since at this moment $F_{,R}/F \ll a^2/k^2$ for all scales of
interest, so $G_{\text{eff}}=G$ in Eq.~\eqref{de1} irrespective of an
actual structure of $F(R)-1$. In turn, if $F_{,R}/F \ge a^2/k^2$,
the $R^2$ correction is negligible for all scales of interest at
the matter-dominated stage. Therefore, this high-$R$ correction to
the model \eqref{model} needed to obtain a viable cosmological
model of the early Universe does not change our results.

\section*{Acknowledgments}

AS acknowledges RESCEU hospitality as a visiting professor. He was also
partially supported by the grant RFBR 08-02-00923 and by the Scientific
Programme ``Astronomy'' of the Russian Academy of Sciences.
This work was supported in part by
JSPS Grant-in-Aid for Scientific Research No.\ 19340054(JY),
JSPS Core-to-Core program  ``International Research Network on Dark Energy'', and
Global COE Program ``the Physical Sciences Frontier'', MEXT, Japan.


\begin{thebibliography}{0}
\bibitem{SS00}
  V.~Sahni and A.~A.~Starobinsky,
  {\it Int.\ J.\ Mod.\ Phys.\ D} {\bf 9}, 373 (2000)
  [arXiv:astro-ph/9904398].
\bibitem{Padmanabhan:2002ji}
  T.~Padmanabhan,
  {\it Phys.\ Rept.}\  {\bf 380}, 235 (2003)
  [arXiv:hep-th/0212290].
\bibitem{Copeland:2006wr}
  E.~J.~Copeland, M.~Sami and S.~Tsujikawa,
  {\it Int.\ J.\ Mod.\ Phys.\  D} {\bf 15}, 1753 (2006)
  [arXiv:hep-th/0603057].
\bibitem{SS06}
  V.~Sahni and A.~A.~Starobinsky,
  {\it Int.\ J.\ Mod.\ Phys.\ D} {\bf 15}, 2105 (2006)
  [arXiv:astro-ph/0610026].
\bibitem{FTH08}
  J.~A.~Frieman, M.~S.~Turner and D.~Huterer,
  {\it Ann.\ Rev.\ Astron.\ Astroph.} {\bf 46}, 385 (2008)
  [arXiv:astro-ph/0803.0982].
\bibitem{Komatsu:2008hk}
  E.~Komatsu {\it et al.}  [WMAP Collaboration],
  {\it Astrophys.\ J.\ Suppl.}\  {\bf 180}, 330 (2009)
  [arXiv:astro-ph/0803.0547].
\bibitem{Yokoyama:2001ez}
  J.~Yokoyama,
  {\it Phys.\ Rev.\ Lett.\ }  {\bf 88},  151302 (2002)
  [arXiv:hep-th/0110137].
\bibitem{SF08}
  T.~P.~Sotiriou and V.~Faraoni
  [arXiv:gr-qc/0805.1726].
\bibitem{S80}
  A.~A.~Starobinsky,
  {\it Phys.\ Lett.\ B} {\bf 91}, 99 (1980).
\bibitem{Hu:2007nk}
  W.~Hu and I.~Sawicki,
  {\it Phys.\ Rev.\  D} {\bf 76}, 064004 (2007)
  [arXiv:astro-ph/0705.1158].
\bibitem{Starobinsky:2007hu}
  A.~A.~Starobinsky,
  {\it JETP Lett.}\  {\bf 86}, 157 (2007)
  [arXiv:astro-ph/0706.2041].
\bibitem{Tsujikawa:2007gd}
  S.~Tsujikawa,
  {\it Phys.\ Rev.\  D} {\bf 76}, 023514 (2007)
  [arXiv:astro-ph/0705.1032].
\bibitem{SHS07}
  Y.-S.~Song, W.~Hu and I.~Sawicki,
  {\it Phys.\ Rev.\ D} {\bf 75}, 044004 (2007)
  [arXiv:astro-ph/0610532].
\bibitem{BEPS00}
  B.~Boisseau, G.~Esposito-Farese, D.~Polarski and
  A.~A.~Starobinsky,
  {\it Phys.\ Rev.\ Lett.} {\bf 85}, 2236 (2000)
  [arXiv:gr-qc/0001066].
\bibitem{SPH07}
  Y.-S.~Song, H.~Peiris and W.~Hu,
  {\it Phys.\ Rev.\ D} {\bf 76}, 063517 (2007)
  [arXiv:astro-ph/0706.2399].
\bibitem{HS07}
  W.~Hu and I.~Sawicki,
  {\it Phys.\ Rev.\ D} {\bf 76}, 104043 (2007)
  [arXiv:astro-ph/0708.1190].
\bibitem{Tsujikawa:2008uc}
  S.~Tsujikawa, K.~Uddin, S.~Mizuno, R.~Tavakol and J.~Yokoyama,
  {\it Phys.\ Rev.\  D} {\bf 77}, 103009  (2008)
  [arXiv:astro-ph/0803.1106].
\bibitem{GMP09}
  R.~Gannouji, B.~Moraes amd D.~Polarski,
  {\it J.\ Cosm.\ Astroph.\ Phys.} {\bf 0902}, 034 (2009)
  [arXiv:astro-ph/0809.3374].
\bibitem{TT08}
  T.~Tatekawa and S.~Tsujikawa,
  {\it J.\ Cosm.\ Astroph.\ Phys.} {\bf 0809}, 009 (2008)
  [arXiv:astro-ph/0807.2017].
\bibitem{O08}
  H. Oyaizu,
  {\it Phys.\ Rev.\ D} {\bf 78}, 123523 (2008)
  [arXiv:astro-ph/0807.2449].
\bibitem{OLH08}
  H. Oyaizu, M.~Lima and W.~Hu,
  {\it Phys.\ Rev.\ D} {\bf 78}, 123524 (2008)
  [arXiv:astro-ph/0807.2462].
\bibitem{SLOH09}
  F.~Schmidt, M.~Lima, H.~Oyaizu and W.~Hu,
  {\it Phys.\ Rev.\ D} {\bf 79}, 083518 (2009)
  [arXiv:astro-ph/0812.0545].
\bibitem{delaCruzDombriz:2008cp}
  A.~de la Cruz-Dombriz, A.~Dobado and A.~L.~Maroto,
  {\it Phys.\ Rev.\  D} {\bf 77}, 123515 (2008)
  [arXiv:astro-ph/0802.2999].
\bibitem{ACD08}
  K.~N.~Ananda, S.~Carloni and P.~K.~S.~Dunsby
  [arXiv:astro-ph/0809.3673].
\bibitem{ACD08a}
  K.~N.~Ananda, S.~Carloni and P.~K.~S.~Dunsby
  [arXiv:astro-ph/0812.2028].
\bibitem{Tegmark:2006az}
  M.~Tegmark {\it et al.}  [SDSS Collaboration],
  {\it Phys.\ Rev.\  D} {\bf 74}, 123507 (2006)
  [arXiv:astro-ph/0608632].
\bibitem{BDK08}
  A.~O.~Barvinsky, C.~Deffayet and A.~Yu.~Kamenshchik,
  {\it J.\ Cosm.\ Astroph.\ Phys.} {\bf 0805}, 020 (2008)
  [arXiv:astro-ph/0801.2063].
\bibitem{F08}
  A.~V.~Frolov,
  {\it Phys.\ Rev.\ Lett.} {\bf 101}, 061103 (2008)
  [arXiv:astro-ph/0803.2500].
\bibitem{D08}
  A.~Dev {\it et al.},
  {\it Phys.\ Rev.\  D} {\bf 78}, 083515 (2008)
  [arXiv:hep-th/0807.3445].
\bibitem{KM09}
  T.~Kobayashi and K. Maeda,
  {\it Phys.\ Rev.\  D} {\bf 79}, 024009 (2009)
  [arXiv:astro-ph/0810.5664].

\end{thebibliography}
\end{document}